# Evaluation of the Central Copper Contact Pin Effect in High-Energy Region in Gamma-ray Spectrometry

E. Uyar[*]

Gazi University, Faculty of Sciences, Department of Physics, Ankara, Turkey

[*]Corresponding author: Esra Uyar
E-mail: esrauyar@gazi.edu.tr
Tel: +90 312 202 12 36
Fax: +90 312 212 22 79

**ABSTRACT**

The detector must be modeled in the most accurate way when Monte Carlo simulation method is used for efficiency calculation in gamma-ray spectrometric studies. This study aims to investigate the effect of the copper contact pin inside the detector on the efficiency of the HPGe detector for high gamma-ray energies. Simulated efficiencies were determined for 6 different energies in the energy range of 1460.8 keV up to 2614.5 keV in point and cylindrical source geometry. According to the modeling using PHITS Monte Carlo simulation code, the presence of copper contact pin at high gamma-ray energies caused a decrease of up to 6% in detector efficiency. It was emphasized that this ignored parameter should be included in the modeling like all other geometric parameters used in detector modeling, by showing the effect on the efficiency.

## 1. Introduction

Gamma-ray spectrometry is one of the most powerful methods used to determine the amounts of radionuclides in any sample. To quantify the radionuclides, it is necessary to know the full-energy peak efficiency of the detector for the energies of interest [1]. Because the number of detected photons is proportional to the concentration of the isotope, a full-energy peak provides all the data required for identifying and measuring a radioactive isotope in a sample. Since it is always necessary for the analysis of a sample unless a standard with the same features is available, the detector efficiency calibration in gamma-ray spectrometry is a highly significant issue. Monte Carlo simulation programs, which have been gaining popularity in many areas over the years, help users significantly, especially in the process of determining efficiency in gamma-ray spectrometry [2]. These programs, which work with an algorithm in which random numbers are generated for random variables, physical experiments are simulated on the computer. In gamma-ray spectrometric studies, high purity germanium (HPGe) detectors can be modeled using these programs and efficiency values can be successfully obtained. In detector modeling, the "quality assurance data sheet" in which many geometric dimension information of the detector is given when the detector is first supplied by the manufacturer, is critical [3]. Because detector simulation is mainly based on information provided by the manufacturer [4]. If the manufacturer gives the detector parameters missing, the accuracy of the simulations is directly affected [5–7]. The accuracy of the simulation results depends on adequate input data and the accuracy of the various approaches applied in the physical model [8]. From the information provided by the manufacturer, for example, dead layer thickness is a time-varying parameter, it is critical to use the existing dead layer thickness determined in the modeling [9, 10]. Otherwise, in the study by Dokania et al. (2014), the physical characteristics of the detector given by the manufacturer were irradiated with the Ge crystal using the radiography method, and they determined that the active detector volume was 20% smaller than the value given by the manufacturer [11]. Therefore, while information is still given about these and similar parameters that need to be tested for accuracy and current status, no manufacturer provides information about the center of the detector is where a copper contact pin is positioned.

The copper contact pin is one of the significant components of the HPGe detector; it is used as both an electrical contact pin and a heat conductor to keep the detector cool (Fig. 1). There are several articles where the presence of this copper contact pin, which reduces the efficiency of the detector, is mentioned, and since its size is not given by the manufacturer, it is included in the modeling with estimated values [12–15].

The research question and purpose of this study is to examine the effect of this copper contact pin, which is located in the middle of the crystal and whose dimensions are not given by any manufacturer, on the detector efficiency at high gamma-ray energies. The necessity of giving the copper contact pin size in the "quality assurance data sheet", which includes the geometric dimension information, is shown with the data obtained from the simulation program.

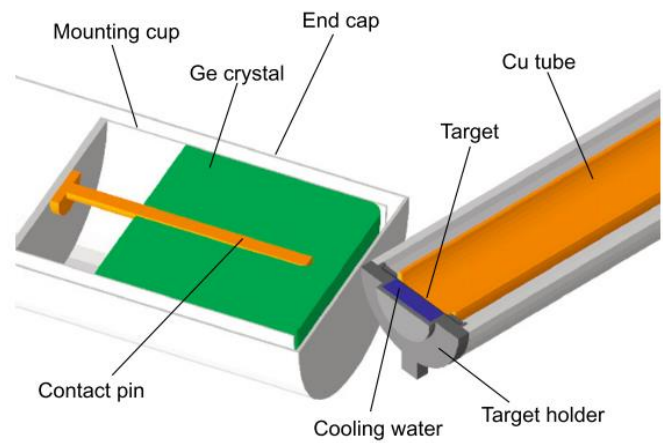

**Fig.1** Contact pin in the setup displaying the germanium detector in the investigation by Carson et al. (2010).

Carson et al. assumed that the contact pin design is proprietary information and is not easily given by the manufacturers, estimating as much as possible the geometry of the contact pin inside the crystal hole from a general drawing provided by the manufacturer, as a solid copper rod with a diameter of 6.9 mm [12]. Östlund et al. used 9.2 mm as the contact pin thickness while modeling the detector in Monte Carlo simulations in their study investigating how the peak-to-valley (PTV) ratio is affected by the detector properties. This value is the diameter of the central hole, so they assumed that the contact pin filled the centre hole of the detector. Because they stated that X-ray and computed tomography could not predict the central electrode diameter [16]. Since the X-ray image did not provide sufficient information to obtain the actual dimensions of the contact pin in Dryak and Kovar, they obtained an image of the copper contact pin by taking the gamma-ray radiography with the Ir-192 gamma source and determined its diameter as 3.7 mm [13].

As can be seen from these limited studies in the literature, no definite interpretation can be made about the dimensions of the contact pin. Therefore, in this study, the detector was modeled in three different ways. It is first modeled without including the contact pin, then with the 3.5 mm contact pin, and finally with the 4.5 mm contact, which is the inner hole diameter of the detector. Also, there is one study in the literature on how this contact pin will have an effect on the detector response and the full energy peak efficiency [17]. In this paper, the effect of the copper contact pin on the detector efficiency was examined in the point source geometry in the energy range of 59.5 keV-1408 keV and it was determined that it changed of up to 1.9% in the efficiency of the detector. The current study investigates the effect of a copper contact pin on the full energy peak efficiency by calculating efficiency values for higher gamma ray energies up to 2614.5 keV in both cylindrical and point source geometry in the high-energy region where the effect is dominant.

## 2. Experimental Details

*2.1 PHITS toolkit for Monte Carlo simulations*

The PHITS Monte Carlo code (version 3.28) was employed to simulate the transport of radiation to create a model for the HPGe detector. In PHITS, source files, binary, data libraries, graphic utility, etc. all contents are fully integrated in one package. The latest version of ENDF (Evaluated Nuclear Data File) and JEFF (Joint Evaluated Fission and Fusion), containing more than 1000 γ-ray spectra, are used as nuclear data libraries. PHITS is a general-purpose Monte Carlo particle transport simulation code that is used in many studies in the fields of accelerator technology, radiotherapy, space radiation, nuclear applications, etc. [18]. It can also be used successfully used in gamma-ray spectrometry to model and respond to HPGe detector [19]. The detector was modeled with PHITS computer code using all data provided by the manufacturer. These data are; detector diameter and length, hole diameter and depth, mount cup length and wall thickness, end cap wall and window thicknesses, hole and outside contact layer (dead layer) thicknesses, etc. The [t-deposit] tally was used to collect the energy (Pulse Height Distribution) deposited in a given region, per emitted gamma particle. This tally provides the energy distribution of the pulses generated in the active germanium crystal. Accordingly, the full energy peak efficiency values used in the study were obtained using the [t-deposit] tally. Since Monte Carlo is a statistical process in which random numbers are used, keeping the number of repetitions as high as possible allows us to obtain more meaningful results. For this reason, one hundred million source particles were used to obtain an uncertainty less than 1% in simulated efficiency in this study.

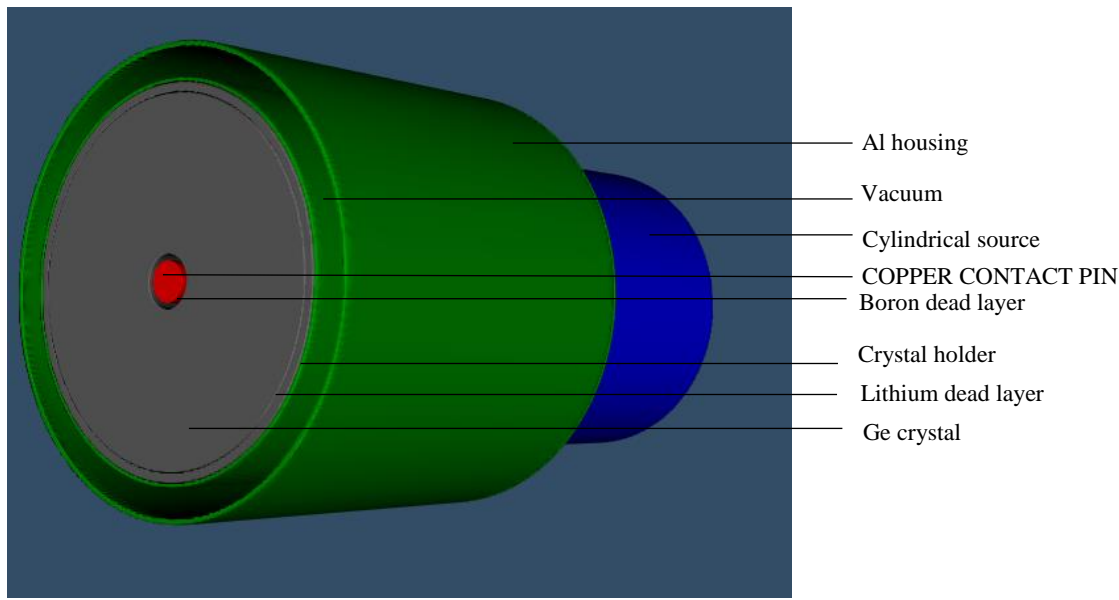

**Fig.2** 3D view of the HPGe detector modeled in PHITS

The position of the copper contact pin in the detector system is shown using the PHIG-3D (PHITS Interactive Geometry viewer in 3D), which reads the PHITS input file and visualizes the geometry in 3D (Fig. 2).

*2.2 Efficiency simulations*

The modeled detector is a p-type coaxial HPGe (PGT IGC50195) with a relative efficiency of 54.7% and a full width at half maximum (FWHM) of 3.8 keV at 1332.5 keV ($^{60}$Co), 1.2 keV at 122.1 keV ($^{57}$Co). Its peak-to-Compton ratio is 67.2:1 at 1332.5 keV ($^{60}$Co). The Ge crystal has a 65.8 mm diameter and a 65.8 mm length. The hole of the detector is 9 mm in diameter and 53 mm depth. The detector with 0.5 mm Al window has a crystal to window distance of 5 mm. The dead layer thickness, which changes over time and significantly affects the full energy peak efficiency, is given by the manufacturer as <1 mm. The dead layer thickness, which was given as <1 mm, was determined as 1.71 mm in our previous study and this value was used in the detector modeling [19].

In the study, efficiency calculations were made for point source geometries counted at certain distances (on the end cap (0 cm), 5 cm, 10 cm, 15 cm and 20 cm) from the detector and cylindrical source geometries counted on the end cap. For this purpose, 1173.2 keV ($f_\gamma$: 99.85%), and 1332.5 keV ($f_\gamma$: 99.98%) peaks of $^{60}$Co with high gamma-ray energy and 1836.1 keV ($f_\gamma$: 99.35%), peaks of $^{88}$Y were used as point sources. In PHITS, the source information is set in the [source] section. The source type is specified with the number s-type=N. Point source modeling was done by choosing s-type=9, which is the source definition for the sphere or spherical surface. Calculations were made considering the 1460.8 keV ($^{40}$K - $f_\gamma$: 10.55%) peak of IAEA-RGK-1, 1764.5 keV ($^{214}$Bi/$^{238}$U - $f_\gamma$: 15.31%), peak of IAEA-RGU-1 and 2614.5 keV ($^{208}$Tl/$^{232}$Th - $f_\gamma$: 99.76%) peak of IAEA-RGTh-1, which are the most calculated/used in environmental radioactivity calculations in cylindrical geometry and have high gamma emission probability ($f_\gamma$). Cylindrical source modeling was done by choosing s-type=1. In s-type=1, the coordinates of the sphere ($x_0$, $y_0$, $z_0$) and its radius, $r_0$, are defined in cylindrical source modeling. The geometric dimensions of the sample container consisting of cylinder acrylic material ($\rho$=1.19 g/cm$^3$) were taken as 5 cm inner height, 3 cm inner radius, 0.15 cm wall thickness, and 0.2 cm bottom thickness. Density and elemental compositions given in Table 1 of IAEA-RGK-1, IAEA-RGU-1 and IAEA-RGTh-1 are defined in the [material] section of PHITS.

**Table 1**
Densities and elemental compositions of the reference materials

| Reference material | Density (g/cm$^3$) | Elemental compositions (%)[*] |
|---|---|---|
| **IAEA-RGU-1** | 1.335 | O: 53.4; Si: 46.4; Al: 0.10 ; U: 0.04; Ca: 0.03<br>Fe: 0.03; Na: 0.02; C: 0.01; Pb: 0.008 ; K: 0.002 |
| **IAEA-RGTh-1** | 1.325 | O: 52.8; Si: 45.6; Y: 0.76; Ca: 0.50; Fe: 0.11; P: 0.11<br>Th: 0.08; K: 0.02; Mg: 0.02; Sr: 0.016; Al: 0.012; Zn: 0.011 |
| **IAEA-RGK-1** | 1.577 | K: 44.8; O: 36.7; S: 18.4 |

[*]Elemental compositions of the reference material derived from XRF data.

In this study, three approaches were made to model the contact pin thickness. Firstly, without the copper contact pin, then by adding a contact pin with a radius of 3.5 mm, and finally for the 4.5 mm value, where the contact pin is assumed to fill the central hole, as in the work of Östlund et al. [16].

## 3. Results and Discussion

In the point source geometry, for 1173.2 keV, 1332.5 keV and 1836.1 keV energies on the detector window (0 cm), at 5 cm, 10 cm, 15 cm and 20 cm distances; in the cylindrical source geometry, efficiency values were obtained for 1460.8 keV, 1764.5 keV and 2614.5 keV energies on the detector window (Table 2). As expected, the efficiency value decreases with the inclusion of the copper contact pin and the increase in its thickness.

**Table 2**

Simulated efficiencies with PHITS for different contact pin values in point and cylindrical source geometry

| Source geometry | Distance | Nuclide | Energy (keV) | Without contact pin | With 3.5 mm contact pin | With 4.5 mm contact pin |
|---|---|---|---|---|---|---|
| Cylindrical | On the end cap | RGK-1 ($^{40}$K) | 1460.8 | 1.158E-02 | 1.119E-02 | 1.099E-02 |
| | | RGU-1 ($^{238}$U) | 1764.5 | 1.033E-02 | 9.973E-03 | 9.807E-03 |
| | | RGTh-1 ($^{232}$Th) | 2614.5 | 7.452E-03 | 7.174E-03 | 7.075E-03 |
| Point | On the end cap (0 cm) | $^{60}$Co | 1173.2 | 4.700E-02 | 4.546E-02 | 4.462E-02 |
| | | $^{60}$Co | 1332.5 | 4.257E-02 | 4.116E-02 | 4.051E-02 |
| | | $^{88}$Y | 1836.1 | 3.289E-02 | 3.174E-02 | 3.113E-02 |
| | 5 cm | $^{60}$Co | 1173.2 | 7.830E-03 | 7.604E-03 | 7.477E-03 |
| | | $^{60}$Co | 1332.5 | 7.135E-03 | 6.918E-03 | 6.810E-03 |
| | | $^{88}$Y | 1836.1 | 5.550E-03 | 5.373E-03 | 5.281E-03 |
| | 10 cm | $^{60}$Co | 1173.2 | 3.024E-03 | 2.943E-03 | 2.894E-03 |
| | | $^{60}$Co | 1332.5 | 2.761E-03 | 2.678E-03 | 2.636E-03 |
| | | $^{88}$Y | 1836.1 | 2.150E-03 | 2.079E-03 | 2.040E-03 |
| | 15 cm | $^{60}$Co | 1173.2 | 1.586E-03 | 1.539E-03 | 1.515E-03 |
| | | $^{60}$Co | 1332.5 | 1.447E-03 | 1.401E-03 | 1.377E-03 |
| | | $^{88}$Y | 1836.1 | 1.146E-03 | 1.108E-03 | 1.088E-03 |
| | 20 cm | $^{60}$Co | 1173.2 | 9.796E-04 | 9.515E-04 | 9.355E-04 |
| | | $^{60}$Co | 1332.5 | 8.928E-04 | 8.643E-04 | 8.463E-04 |
| | | $^{88}$Y | 1836.1 | 7.030E-04 | 6.784E-04 | 6.638E-04 |

There is true coincidence summing effect in gamma-ray spectrometry, especially in low source-to-detector distance and radionuclides with complex decay scheme that emit more than one gamma-ray [20]. In this study, all radionuclides except $^{40}$K have this effect. The true coincidence summing factor is an important correction factor affecting the full energy peak efficiency. However, since the ratios of the full energy peak efficiency values obtained in the study (without contact pin/with 3.5 mm contact pin and without contact pin/with 4.5 mm contact pin) are used, this factor will not have an effect on the results.

The effect of the contact pin was investigated by calculating the % difference between the efficiency values calculated without the contact pin and the efficiency values determined with the 3.5 mm and 4.5 mm contact pin. Accordingly, this effect is given in Fig.3 in point source geometry and in Fig.4 in cylindrical source geometry. In the existence of a copper contact pin with a radius of 3.5 mm, the percent difference in detector efficiency, ie reduction values; 2.7-3.3% at 1173.2 keV; 3.0-3.3% at 1332.5 keV; 3.2-3.5% at 1836.1 keV. In the existence of a copper contact pin with a radius of 4.5 mm, these values are; 4.3-5.1% at 1173.2 keV; 4.5-5.2% at 1332.5 keV; 4.9-5.6% at 1836.1 keV (Fig.3).

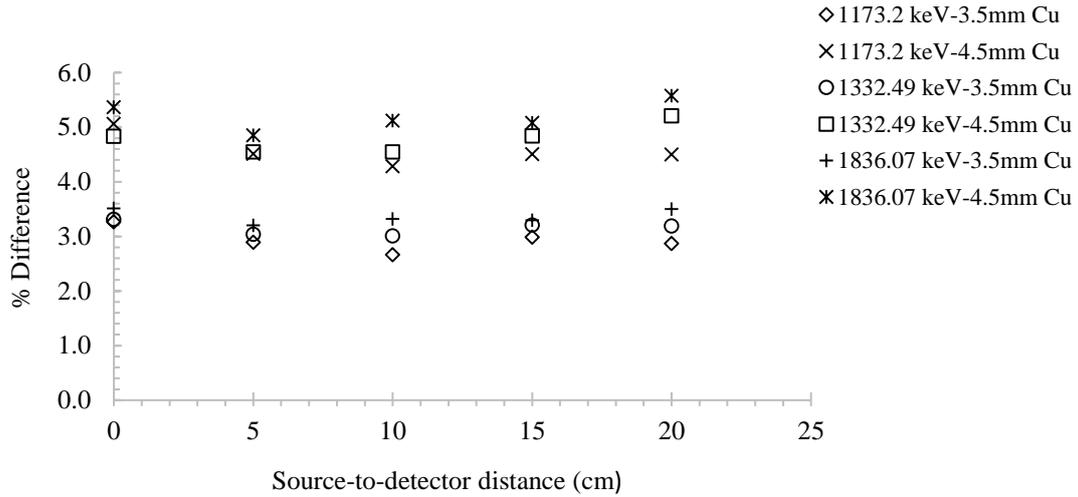

**Fig.3** The difference between the efficiency values determined with 3.5 mm and 4.5 mm contact pin at different source-to-detector distances and the efficiency value determined without a contact pin (point source geometry)

In the presence of a 3.5 mm radius copper contact pin, the percent difference in the detector efficiency; 3.4% at 1460.8 keV; 3.4% at 1764.5 keV; 3.7% at 2614.5 keV. In the presence of a copper contact pin with a radius of 4.5 mm, these values are; 5.1% at 1460.8 keV; 5.0% at 1764.5 keV; 5.1% at 2614.5 keV (Fig.4). Similar differences at 1460.8 keV, 1764.5 keV and 2614.5 keV show that the effect does not increase with further increase in energy. Therefore, the biggest difference that can be created in the detector efficiency because of interactions that may occur at high energy is at these levels.

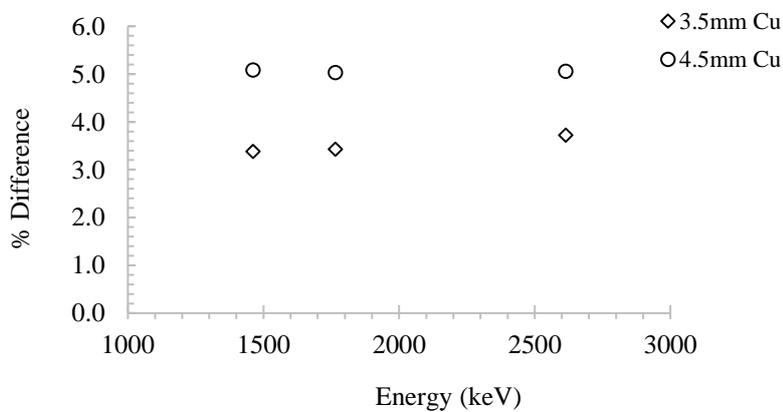

**Fig.4** The difference between the efficiency values determined with 3.5 mm and 4.5 mm copper at a fixed source-to-detector distance and the efficiency value determined without copper (cylindrical source geometry)

Compared with the 1836.1 keV peak in the point source geometry, the 1764.5 keV peak in the cylindrical source geometry at close energies, the efficiency change in both energies is ≈ 3.5% at the 3.5 mm contact pin; in 4.5 mm contact pin, it is seen that ≈ 5%. Consequently, it can be said that the copper contact pin is independent of the source geometry.

## 4. Conclusions

The manufacturer-provided parameters are critical in simulation calculations where HPGe detectors are modeled. The design of the contact pin is proprietary information and cannot be easily obtained from the manufacturers. While many parameters affecting the efficiency of the detector are investigated in detail in the literature, the effect of the copper contact pin is ignored. The effect of copper contact pin thickness on the detector efficiency in the 1460.8 keV-2614.5 keV energy range was investigated and the decrease in the efficiency increases to approximately 4% when the contact pin thickness is modeled as 3.5 mm, and 6% for 4.5 mm. Since photons with low-energy are absorbed because of photoelectric interaction before they reach that region, only high gamma energy peaks are considered in the study. The peaks in the high-energy region interact with the contact pin, which is the material in the interior of the crystal, due to the Compton scattering and pair production events dominant in this region, decreasing the efficiency values.

There are many sources of uncertainty (such as measurement geometry, decay graph and input data) in determining the efficiency value by simulation in the activity concentration calculation using gamma-ray spectrometry. Therefore, not including a copper contact pin in addition to these uncertainty sources brings extra uncertainty by taking it away from the true value. When modeling the detector in general-purpose Monte Carlo programs, the copper contact pin should be included in the coding so that its center is in the middle of the hole. In dedicated-purpose programs, the developers of the program should add this parameter as a contact pin to the part where the detector parameters are defined. It is also critical to determine the efficiency values to be obtained from the simulation to be used in the activity concentration calculations of the samples containing natural radionuclides such as U, Th, K, which are also considered in this study.

**Acknowledgments**

The author is thankful to Prof. Emeritus Octavian Sima, University of Bucharest, Physics Department, for his helpful information on the subject.